\documentclass[twocolumn,twoside,slac_two]{revtex4}
\usepackage{graphicx}
\usepackage{fancyhdr}
\begin{document}

\title{Communicating Science to the Media}

%

\author{Kathryn Grim}
\affiliation{Fermi National Accelerator Laboratory, Box 500, Batavia, IL 60510, USA}

\begin{abstract}
If you watched the Daily Show's feature on the Large Hadron Collider in May, you will remember the look on John Ellis' face when John Oliver asked him, ``Evilgeniussayswhat?" It's impossible to anticipate everything a reporter will ask you, but this media training will teach you how to prepare for interviews with print, television or radio journalists and increase your understanding of how to communicate effectively through the media. 

Kathryn Grim of the Fermilab Office of Communication will provide a general overview of how the media works, what makes news and why. She will explain the importance of key messages, analogies and sound bites and conduct some exercises on how to craft them. By the end of the session, you will understand the importance of preparing for media interactions, gain clarity on how to successfully present your work, and understand where to go for assistance in dealing with the media, just in case the Daily Show ever comes knocking on your door.

\end{abstract}

\maketitle

\thispagestyle{fancy}

This media training is meant to give participants a better idea of how the media works, what reporters are looking for, and how to prepare for interactions with reporters. Participants should learn how to define key messages, avoid jargon and provide good sound bites.

A reporter wants to write or produce a piece that will interest readers, viewers or listeners. The seven qualities that make a story interesting are its impact, immediacy, proximity, prominence, novelty, conflict and emotion \cite{Harrower}.

An audience will be interested in a story with {\em impact}, one that speaks about something that will have an effect on them or their pocketbooks. A story is more newsworthy if it has {\em immediacy}, if it just happened or is about to happen. Reporters often search for stories with {\em proximity} because events that occur close to home will matter more to readers. Audience members also care about stories involving people with {\em prominence}, celebrities or other well-known public figures.

Articles can draw readers in with {\em novelty}; if a dog bites a man, that's nowhere near as interesting as a man biting a dog. Finally, audience members are interested in stories with {\em conflict} and those that elicit an {\em emotional} response. If you can highlight one or more of these qualities in the story you'd like to tell the public, reporters will be happier to pick it up and their audiences will be more interested.

The first and most important way to prepare for an interview is to develop your key message. This should be one or two sentences that answer the questions: What is the main point of your story and why is it important?

You should prepare to give more details to flesh out your story, but be sure to begin the interview with your main point. It may be what you hope to discover. It may be that your experiment is the largest of its kind. It may be that you are using a novel method to make your discovery. No matter what your main point is, be sure to explain why people should care. What is your goal? How will this affect others?

Your key message should fit nicely into a sound bite. A sound bite could be a direct quotation in a printed story or a clip played on the radio or television. Reporters use direct quotations rather than paraphrasing your words when your words are specific, vivid, descriptive, or show your personality. So if you speak only in bland phrases full of jargon, you're going to give the reporter a hard time finding a way to make your subject look interesting.

 You should prepare a few sound bites before an interview. Make sure they are one or two short sentences long, easy to remember and take 10 seconds or less to say.

When preparing for an interview, you should be ready to answer seven basic questions about your research. Who is involved? What is it? When does it take place? Where does it take place? Why are you doing it? How are you doing it? And why should we care?

But there are other possibilities for which to prepare, such as: Is this dangerous? What could go wrong? How much does this cost? Wouldn't the money be better spent on finding a cure for cancer?

These questions should not bother you. A reporter's job is to ask the questions that his or her audience members would want to ask. Taxpayers want to know how their dollars are being spent. Readers who have a family member with a debilitating disease want to know if our country's great scientific minds are wasting their time when they're studying something else. These people and their concerns matter. The better you can explain to them what you're doing and how it will benefit them, the more support your research will have.

The key is to be understandable and positive. 

Part of being understandable is avoiding jargon -  words that make sense to a physicist but sound like gibberish to the rest of the public. There are two ways to deal with jargon. Either use it and explain it with a definition or an analogy or don't use it at all.

To find examples of jargon explained, try reading ``Explain it in 60 Seconds" \cite{symmetry} features from symmetry magazine. Here is how the magazine explained antimatter:

``Consider this analogy: dig a hole, and make a hill with the earth you've excavated. Hole and hill have equal but opposite characteristics - the volume of the earth in the hill, and that of the hole where the earth was removed. For particles, properties like electrical charge are opposite to their antiparticles - one positive, one negative.''

``Also, antimatter will annihilate its matter counterpart in a burst of energy, just like the hill will fill the hole, leaving neither."

To avoid using jargon, think of another way to say what you'd like to say. To avoid having to explain what a nanometer is while describing the diameter of atom, describe how many atoms fit on a pinhead or compare the size of an atom to the diameter of a human hair.

And stay positive. Do not repeat a negative idea, even to deny it. If a reporter asks you, for example, if your experiment is going to cause a black hole, it is better to say, ``The experiment is perfectly safe," than, ``The experiment will not cause a black hole." Someone watching you say the latter on the news is going to wonder why you need to talk about the possibility of black holes at all.

Most reporters would not use such a quote for the sole purpose of making you look suspicious. However, a good reporter cannot take the statement, ``The experiment will not cause a black hole" to mean ``The experiment is perfectly safe." You may say there will be no black hole, but that does not preclude the possibility that your experiment will cause a huge ball of fire instead.  You need to put things in a positive light; the reporter cannot do it for you.
So before the interview, think of difficult questions you might have to answer, and come up with positive ways to answer them.

You should also think of questions you would like to ask the reporter. Ask what kind of ground you will be covering in the interview. What kind of message is the reporter looking for? How will it be used? With what other material? Whom else will be interviewed for this piece? If the piece is for radio or television, ask if the interview will be live or recorded. How long will the interview last? Where and how will it take place?

If you are cold-called and asked to do a phone interview, tell the reporter that you are busy and set up a good time to call back. This will give you time to prepare. But realize that reporters are usually working on a tight deadline, so if you make them wait too long, they will have to find someone else to interview.

Talking on the phone can feel comfortable and informal, maybe a little too much so. To avoid saying something you will regret, imagine that someone whose opinion you care about is standing behind you as you talk.

Conversely, a live interview can make you nervous. If so, try to think of it as a social chat. If you are preparing for a television or radio interview, arrive early. If the interview is pre-recorded, feel free to ask to try to explain something again if you do not think you did well on your first try. During a radio interview, talking with your hands can make you sound more natural.

In preparing for a television interview, check your appearance. Dress quietly, without bold patterns or dangly earrings. Wear summer-weight clothing, as studio lights are hot. Avoid wearing tinted lenses. If someone at the studio offers to change your clothes or makeup, trust him or her. 

During the interview, sit forward rather than leaning back, which will make you seem disengaged from the conversation. Do not cross or splay your legs. Look at the interviewer, not the camera, and use normal body language. If you are unsure where to look, ask.

During any interview state the most important information first and give the background second. Keep your responses brief but long enough to give the reporter quotes to use. Stick to your key messages, repeating if necessary. Mention the subject you're discussing by name - rather than saying ``it" or ``this" - several times during the interview to create better sound bites, ones that need less or no introduction. 

Do not overestimate a reporter's knowledge of your subject. Give background and set the record straight if the reporter seems to be asking a question based on incorrect information. Be sure to identify whether something is a fact or your opinion. If you do not understand a question, ask for clarification rather than risking giving a confusing answer. If you do not know the answer, tell the reporter you will get back to him or her; inventing something off the top of your head will come back to bite you!

Just as you shouldn't make things up, you should also correct reporters when they are wrong. They will appreciate it. They do not want to look bad any more than you do, and mistakes in their stories reflect badly on them. But make sure to do so without being argumentative. Realize that the reporter is the one who's going to produce the story. You do not want to make yourself look bad with combative quotes. If you are on a live program, realize that the audience is loyal to the reporter. If you try to make him or her look bad, it will wind up reflecting poorly on you.

Instead, be enthusiastic about your research. Let people know what interests you. Your excitement could be infectious and will give you better quotes.

After the interview, be sure the reporter knows how to spell your name and what to give as your title or position. Ask for a copy of the final product or to know when the piece will air. Ask for feedback so that you can be better prepared for your next interview. And thank the reporter for his or her time and interest. With any luck, you will be able to establish a professional relationship, and the reporter will feel comfortable using you as a source in the future.

Interviewing is a skill like any other; in order to improve, you need to practice. Explain your key messages to non-physicists. Ask non-physicists to interview you. 

Watch, read or listen to reports on unfamiliar topics. Think about what interests you and what you remember. Try to listen to yourself that way.

Make sure to work with your university or institution's public affairs office. They can offer advice, answer questions and serve as an excellent test audience. You may also contact me, Kathryn Grim, by email at kgrim@fnal.gov.

\bigskip 

\end{document}